\documentclass[12pt]{article}
\usepackage{amsmath,amssymb}
\usepackage{bm}
\usepackage{color}
\usepackage{graphicx}
\allowdisplaybreaks

\newcommand{\bea}{\begin{eqnarray}}
\newcommand{\eea}{\end{eqnarray}}

\newcommand{\bp}{\textbf{p}}
\newcommand{\bq}{\textbf{q}}
\newcommand{\bk}{\textbf{k}}

\newcommand{\nn}{\nonumber}

\begin{document}
\title{
\begin{flushright}
\ \\*[-80pt] 
\begin{minipage}{0.2\linewidth}
\normalsize
HUPD-2204 \\*[5pt]
\end{minipage}
\end{flushright}
{ \bf 
Time evolution of the lepton number of Majorana neutrinos in the Schr\"{o}dinger 
 picture  versus Heisenberg picture
\\*[5pt]}}
\author{\normalsize
\centerline{
Nicholas J. Benoit$^{1}$\footnote{E-mail address: d195016@hiroshima-u.ac.jp},
Yuta Kawamura$^{1}$\footnote{E-mail address: 
kawamura1994phy@gmail.com},
Takuya Morozumi$^{2,3}$\footnote{E-mail address: morozumi@hiroshima-u.ac.jp
}
} \\ \normalsize
\\*[10pt]
\centerline{
\begin{minipage}{\linewidth}
\begin{center}
$^1${\it \small
Graduate School of Science, Hiroshima University, 
Higashi-Hiroshima 739-8526, Japan} \\*[5pt]
$^2${\it \small
Physics Program, Graduate School of Advanced Science and Engineering, \\
Hiroshima University, Higashi-Hiroshima 739-8526, Japan} \\*[5pt]
$^3${\it \small 
Core of Research for the Energetic Universe, Hiroshima University, \\
Higashi-Hiroshima 739-8526, Japan} 
\\*[5pt]
\end{center}
\end{minipage}}
\\*[50pt]}
\date{
\centerline{\small \bf Abstract}
\begin{minipage}{0.9\linewidth}
\medskip
\small
In this paper, we study the time evolution of the expectation value of Majorana neutrino
with the Schr\"{o}dinger picture.  The operators with the definite lepton number and  operators with the definite mass
are related to each other by a Bogolyubov  transformation.  Then the vacuum with the null lepton number is also related to the vacuum for the massive operator 
 and it is written by the superposition of the vacuum for massive field and  Majorana pairs condensed  states. 
We choose the state with a definite lepton number $L$ $=1$ and the momentum ${\bf q}\ne 0$ as an initial state. By writing the state in terms of the superposition of energy eigenstates, we are able to    
study the time evolution of the state  in the Schr\"{o}dinger picture.  The expectation value of lepton number operator is computed and it reproduces
the same result as that obtained in the corresponding Heisenberg operator.   

\end{minipage}
}
\begin{titlepage}
\maketitle
\thispagestyle{empty}
\end{titlepage}
\section{Introduction}
In the previous paper \cite{Adam:2021qiq},  we have studied the time evolution of the lepton number carried by Majorana neutrinos in the Heisenberg picture.
In the present paper, we investigate the same problem in the 
the Schr\"{o}dinger picture.  
One can compute the expectation value
of the physical observables at arbitrary time in both  Heisenberg picture and 
Schr\"{o}dinger
 picture and they are equivalent in this aspect. However,
in order to find how the state evolves  with respect to  time, it is desirable to formulate the previous framework in the Schr\"{o}dinger picture.  

The operator relation between  neutrino/anti-neutrino and Majorana neutrino obtained 
in \cite{Adam:2021qiq} is expressed  as a Bogolyubov transformation
\cite{Bogolyubov:1958km}.
Then the two vacua,  one which is annihilated by the operators assigned with definite lepton number
 and the other annihilated by the operators with definite mass,  are related by the transformation.
Using the relation, the initial state with the definite lepton number can be expressed
in terms of the operators with the definite mass applied on the vacuum  
 annihilated by the operators with definite mass.  Then the time evolution of the 
initial state can be derived and the expectation value of the lepton number is also
obtained.

The  Bogolyubov transformation is applied to the oscillation phenomena \cite{Blasone:1995zc} and \cite{Tureanu:2018phm}.
The Majorana neutrino as a  Bogolyubov quasi-particle is also suggested in the literature
\cite{Fujikawa:2018rci}. 

The paper is organized as follows.
In section 2, we introduce the  Bogolyubov  transformation to relate the operator for mass eigenstate to the operator with the definite lepton number.  In section 3, the initial
state is given.  The time evolution of the state and the expectation value of lepton number are 
obtained in section 4.
The numerical calculation is also shown. 
Section 5 is devoted to the conclusion and the derivation of 
the  Bogolyubov transformation and the initial state are given in appendixes.  

\section{Continuity condition and The  Bogolyubov  Transformation}
In the previous paper \cite{Adam:2021qiq}, we consider the situation that the Majorana mass term is turned on at $t=0$ by a time step function. The following Lagrangian corresponds to
that situation, but for the single
flavor case,
\bea
{\cal L}=\overline{\nu_L} i \partial_\mu \gamma^\mu \nu_L -\theta(t)\frac{m}{2} \{ \overline{(\nu_L)^c}{\nu_L} +h.c. \} .
\eea
Then  the equation of motion leads to a continuity condition at $t=\pm \epsilon$,
\bea
 \nu_L(t=-\epsilon, {\bf x}) =P_L \psi_M(t=+\epsilon, {\bf x}),
\label{eq:continuity}
\eea
where $P_L$ denotes the left-handed projection $\frac{1-\gamma_5}{2}$. The left-hand side of Eq.(\ref{eq:continuity}) is expanded by  on-shell massless spinors.  Their 
coefficients are the annihilation operator denoted by $a(\bf p)$ for a neutrino 
and the creation operator for an anti-neutrino denoted by $b^\dagger(\bf p)$. The right-hand side is expanded by  massive on-shell spinors. Their coefficients correspond 
to annihilation and creation operators for massive Majorana field
denoted by $a^{(\dagger)}_M({\bf p}, \lambda)$ where $\lambda$ denotes the helicities $\lambda=\pm$. 
The non-zero momentum ${\bf p}$ can be always split into two regions.  A  hemisphere  region is called as  $A$ region and the  other  hemisphere region is called as $\bar
{A}$. (See for the details in \cite{Adam:2021vbl} .)  The continuity condition is transfered to the following relations between the operators, 
  \begin{eqnarray}
    \frac{a (\bp)}{\sqrt{2 p }} & = & \frac{\sqrt{E_p + p}}{2 E_p} \left( a_M
    (\bp, -) + \frac{i m}{E_p + p } a^{\dagger}_M (-\bp, -) \right), \label{relation1}
 \\
    \frac{a (- \bp)}{\sqrt{2 p }} & = & \frac{\sqrt{E_p + p}}{2 E_p} \left(
    a_M (- \bp, -) - \frac{i m}{E_p + p } a^{\dagger}_M (\bp, -) \right),
 \label{relation2} \\
    \frac{b (\bp)}{\sqrt{2 p }} & = & \frac{\sqrt{E_p + p}}{2 E_p} \left( a_M
    (\bp, +) + \frac{i m}{E_p +p } a^{\dagger}_M (- \bp, +) \right),
  \label{relation3} \\
    \frac{b (-\bp)}{\sqrt{2 p }} & = & \frac{\sqrt{E_p + p}}{2 E_p} \left(
    a_M (- \bp, +) - \frac{i m}{E_p + p} a^{\dagger}_M (\bp, +) \right),
\label{relation4}
\end{eqnarray}
where $p=|\bp|$ , $E_p=\sqrt{p^2+m^2} $ and $\bp \in A$.  
To further simplify  relations in Eq.(\ref{eq:continuity}),
 we  introduce the momentum dependent angle $\phi_p$
, which is related to the velocity  $ v_p$ as $v_p =\frac{p}{E_p}= \cos 2 \phi_p$. 
Then we rewrite the operator relations in Eqs.(\ref{relation1}-\ref{relation4}),
  \begin{equation}
    \left(\begin{array}{c}
      a (\bp)\\
      a^{\dagger} (-\bp)
    \end{array}\right) = \sqrt{\cos 2 \phi_p} \left(\begin{array}{cc}
      \cos \phi_p & i \sin \phi_p\\
      i \sin \phi_p & \cos \phi_p
    \end{array}\right) \left(\begin{array}{c}
      a_M (\bp, -)\\
      a_M^{\dagger} (- \bp, -)
    \end{array}\right), \label{aoperators}
  \end{equation} 
 \begin{equation}
    \left(\begin{array}{c}
      b (\bp)\\
      b^{\dagger} (-\bp)
    \end{array}\right) = \sqrt{\cos 2 \phi_p} \left(\begin{array}{cc}
      \cos \phi_p & i \sin \phi_p\\
      i \sin \phi_p & \cos \phi_p
    \end{array}\right) \left(\begin{array}{c}
      a_M (\bp, +)\\
      a_M^{\dagger} (-\bp, +)
    \end{array}\right). \label{boperators}
  \end{equation}
The above relations can be expressed  by using the transformation 
$T(\phi_p) $ defined below,
\bea
T(\phi_p)&=&\exp{[i \phi_p (g_{p,+}+g_{p,-})]} , \label{Transform}\\
g_{p, \lambda} & \equiv & \alpha_M (\bp, \lambda) \alpha_M (- \bp, \lambda) +
\alpha^{\dagger}_M (-\bp, \lambda) \alpha^{\dagger}_M (\bp, \lambda),
\label{eq:Tphip}
\eea
where $g_{p,\pm}$ denote the generators of the transformation.  Note that we have 
 introduced the dimensionless operators, 
\bea
\alpha_M (\bp, \lambda) & \equiv & \frac{a_M (\bp, \lambda)}{\sqrt{2 E_p (2
  \pi)^3 \delta^{(3)} (0)}}.
\eea
They satisfy the anti-commutation relations
for $\bp=\bq$,
\bea
\{\alpha_M (\bp, \lambda), \alpha^\dagger_M (\bq, \lambda') \} =\delta_{\lambda \lambda^\prime}.
\eea
Using the transformation $T(\phi_p)$ in Eq.(\ref{Transform}), one can rewrite the relations in Eqs.(\ref{aoperators}-\ref{boperators}) between massless operators and massive ones   as,
\bea
a(\bp) &=&\sqrt{\cos 2\phi_p}  T(\phi_p) a_M(\bp, -) T^{-1}(\phi_p), \nn \\
          &=&\sqrt{\cos 2\phi_p}  \left(\cos \phi_p a_M(\bp, -)+ i \sin \phi_p 
a_M^\dagger(-\bp, -) \right), 
\label{eq:Bogo1} \\
a(-\bp)&=&\sqrt{\cos 2\phi_p} T(\phi_p) a_M(-\bp, -) T^{-1}(\phi_p), \nn \\
  &=&\sqrt{\cos 2\phi_p}  \left(\cos \phi_p a_M(-\bp, -)- i \sin \phi_p 
a_M^\dagger (\bp, -) \right), 
\label{eq:Bogo2} \\
b(\bp)&=& \sqrt{\cos 2\phi_p}  T(\phi_p) a_M(\bp, +) T^{-1}(\phi_p), \nn \\
&=&\sqrt{\cos 2\phi_p}  \left(\cos \phi_p a_M(\bp, +)+ i \sin \phi_p 
a_M^\dagger(-\bp, +) \right), 
\label{eq:Bogo3} \\
b(-\bp)&=& \sqrt{\cos 2\phi_p} T(\phi_p) a_M(-\bp, +) T^{-1}(\phi_p), \nn \\
&=& \sqrt{\cos 2\phi_p}  \left(\cos \phi_p a_M(-\bp, +)- i \sin \phi_p 
a_M^\dagger (\bp, +) \right).
\label{eq:Bogo4}
\eea
The derivation of the relations in Eqs.(\ref{eq:Bogo1}-\ref{eq:Bogo4}) are shown in the appendix A. The matrix $T$ denotes the  Bogolyubov  transformation
\cite{Bogolyubov:1958km}.
\subsection{ Relation between two vacua from the  Bogolyubov  transformation}
Since there are two sets of operators, $(a(\bp), b(\bp))$ and $a_M(\bp, \lambda=\pm1 )$, one can also define two different vacua.  The vacuum denoted by 
$|0 \rangle $
is annihilated by $a(\pm \bp)$ and $b(\pm \bp)$,
\bea
a(\pm \bp) |0 \rangle =b(\pm \bp) |0 \rangle=0,
\label{vacuum1}
\eea
where  $\bp \in A$ . 
Eq.(\ref{vacuum1}) can be translated as,
\bea
T(\phi_p) a_M(\pm \bp, \lambda=\pm1) T^{-1}(\phi_p) |0 \rangle =0.
\eea
The vacuum $|0_M \rangle$ that is annihilated by the operator $a_M(\pm \bp, \lambda=\pm)$ 
satisfies,
\bea
a_M(\pm \bp, \lambda=\pm)|0_M \rangle=0. 
\eea
The two vacua are related to each other as,
\bea
&& |0 \rangle = T  |0_M \rangle , \quad   T=\prod_{\bp \in A} T(\phi_p).
\eea
Below we explicitly construct the vacuum $|0 \rangle$ by applying the  Bogolyubov  transformation $T$ on the vacuum $|0_M\rangle $. 
\bea
|0 \rangle= T|0_M \rangle=\prod_{\bp \in A} \{ \cos^2\phi_p-\sin^2\phi_p 
B^\dagger_{M +}(\bp)  B^\dagger_{M -}(\bp) 
 +i \sin \phi_p \cos \phi_p  \sum_{\lambda=\pm} B^\dagger_{M \lambda}(\bp) 
 \} |0_M \rangle,
\label{eq:Tp}
\eea
where $B^\dagger_{M \lambda}(\bp)$ is defined as, 
\bea
B^\dagger_{M \lambda}(\bp)=\alpha^\dagger_M(-\bp,\lambda) \alpha^\dagger_M(\bp,\lambda).
\eea 
The derivation can be found in Eq.(\ref{eq:ATp}) in the appendix.
This bosonic operator creates two Majorana particles with opposite momenta. 
We note that there are states of two pairs and one pair of massive Majorana neutrinos  in  superpositions.
The norm of these states are given by,
\bea
&& \langle  0_M|  B_{M -}(\bp) B_{M +}(\bp)  B^\dagger_{M +}(\bp)  B^\dagger_{M -}(\bp)  |0_M \rangle=1, \\
 && \langle  0_M|  (\sum_{\lambda'=\pm} B_{M \lambda'}(\bp) ) ( \sum_{\lambda=\pm} B^\dagger_{M \lambda}(\bp) )
 |0_M \rangle=2. 
\eea
\section{Construction of a one particle state with a definite lepton number}
We are ready to build an initial state with a definite lepton number as follows, 
\bea
|\Psi(0) \rangle &=&  N_q a^\dagger(\bq) |0\rangle \nn \\
&=& N_q  \sqrt{\cos 2\phi_q} (\cos \phi_q a^\dagger_M(\bq,-) - i 
\sin \phi_q a_M(-\bq,-) )  |0\rangle  \nn \\
&=& \Bigl{[} \prod_{\bp \ne \bq \in A} 
\{ \cos^2\phi_p-\sin^2\phi_p 
B^\dagger_{M +}(\bp)  B^\dagger_{M -}(\bp) 
 +i \sin \phi_p \cos \phi_p  \sum_{\lambda=\pm} B^\dagger_{M \lambda}(\bp) 
 \}  \Bigr{]}   \nn \\
&&\quad \times  N_q  \sqrt{\cos 2\phi_q}\left( \cos \phi_q + i  \sin \phi_q B^\dagger_{M+}(\bq) \right)
a^\dagger_M(\bq,-) |0_M \rangle,
\label{eq:initial}
\eea
where $N_q$ is a normalization factor given by,
\bea
N_q=\sqrt{\frac{1}{2 q (2\pi)^3 \delta^{3}(0)}}.
\eea 
To derive Eq.(\ref{eq:initial}), we use Eq.(\ref{eq:Tp}) and  Eq.(\ref{eq:B1}).
\section{Time evolution of the initial state and the expectation value for lepton number}
\begin{figure}[htbp]
\begin{center}
\includegraphics[width=0.6\linewidth]{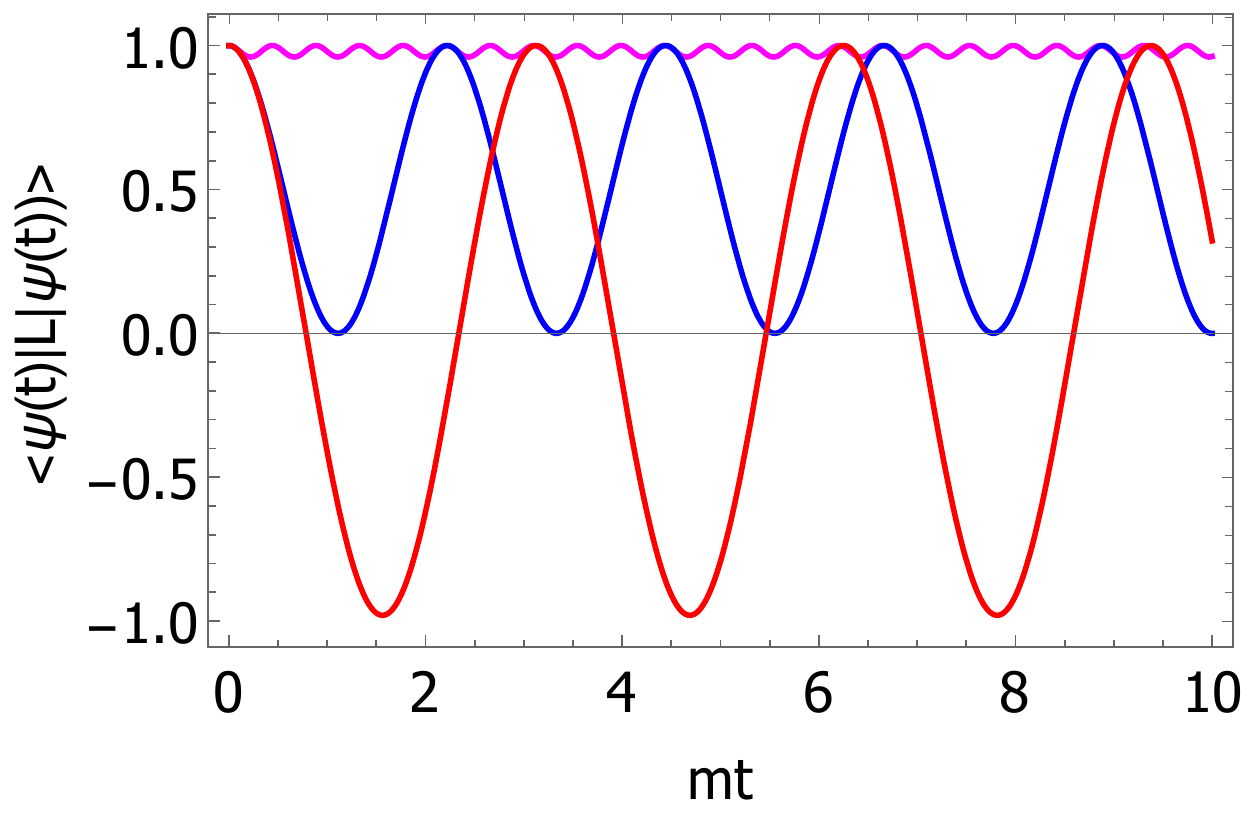}
\caption{ The expectation value of the lepton number is plotted versus $m t$ where $m$ denotes the neutrino mass.
The pink colored line shows the relativistic case with $v=0.99$, The blue colored line shows the case with $v=\frac{1}{\sqrt{2}}$. The red colored line
shows the non-relativistic case with $v=0.1$.}  
\label{fig1}
\end{center}
\end{figure}

The state evolved from the initial state in  Eq.(\ref{eq:initial}) is obtained;
\bea
|\Psi (t) \rangle 
&=&  \prod_{\bk (\ne \bq) \in A}  \{ \cos^2\phi_k-\sin^2\phi_k e^{-4 i E_k t} 
B^\dagger_{M +}(\bk)  B^\dagger_{M -}(\bk) 
 +i \sin \phi_k \cos \phi_k    e^{-i 2 E_k t} 
\sum_{\lambda=\pm 1} B^\dagger_{M \lambda}(\bk) 
 \} \nn \\
&& \times N_q  \sqrt{\cos 2\phi_q} e^{-i E_q t} \left( \cos \phi_q + i  \sin \phi_q  e^{-2 i E_q t} B^\dagger_{M+}(\bq) \right)
a^\dagger_M(\bq,-)  |0_M \rangle.
\label{eq:statet}
\eea
We compute the expectation value of lepton number with the state at $t$ defined in Eq.(\ref{eq:statet}), 
\bea
\langle \Psi(t)|L|\Psi(t) \rangle, 
\label{eq:expextationvalue}
\eea
where the lepton number is originally written in terms of the operator associated with operators for massless neutrinos.
\bea
L&=&\int^\prime  \frac{ d^3p}{(2\pi)^3 2|\bp|} (a^\dagger (\bp) a(\bp) -b^\dagger(\bp) b(\bp)). \nn \\
&=& \int_{\bp \in A }  \frac{ V d^3p}{(2\pi)^3 } l_\bp,
\label{eq:leptonN}
\eea
where $V=(2\pi)^3 \delta^{(3)}(\bp=0)$ and $l_\bp$ is the contribution to the lepton number from the momentum $\bp$ sector 
and is rewritten in terms of the operators for massive Majorana particle.
\bea
l_\bp&=&\frac{|\bp|}{E_p} (n(\bp,-) +n(-\bp,-)-n(\bp,+)-n(-\bp,+) ) \nn \\
& -& \frac{i m}{E_p}  (B_{M-}^\dagger(\bp)-B_{M+}^\dagger(\bp)+B_{M+}(\bp)-B_{M-}(\bp) ),\\
n(\bp,\lambda)&=&\alpha^\dagger_M(\bp,\lambda) \alpha_M(\bp,\lambda).
\eea
The expectation value of $l_\bq$ for the state $|\Psi(t) \rangle $ is given by
\bea
&& \langle \Psi(t) |l_\bq|\Psi(t) \rangle\nn \\
&&=  \prod_{\bk (\ne \bq) \in A}  \langle 0_M | 
 \{ \cos^2\phi_k-\sin^2\phi_k e^{4 i E_k t} 
  B_{M -}(\bk) B_{M +}(\bk)
 -i \sin \phi_k \cos \phi_k    e^{i 2 E_k t} \sum_{\lambda=\pm 1} B_{M \lambda}(\bk) 
 \} \nn \\
&&  \{ \cos^2\phi_k-\sin^2\phi_k e^{-4 i E_k t} 
B^\dagger_{M +}(\bk)  B^\dagger_{M -}(\bk) 
 +i \sin \phi_k \cos \phi_k    e^{-i 2 E_k t} 
\sum_{\lambda=\pm 1} B^\dagger_{M \lambda}(\bk) 
 \}  |0_M \rangle  N_q^2  \cos 2\phi_q  \times
\nn \\
&& \langle 0_M |    a_M(\bq,-) ( \cos \phi_q - i  \sin \phi_q  e^{2 i E_q t} B_{M+}(\bq) ) 
l_\bq 
 ( \cos \phi_q + i  \sin \phi_q  e^{-2 i E_q t} B^\dagger_{M+}(\bq) )
a^\dagger_M(\bq,-)  |0_M \rangle \nn \\
&&= N_q^2  \cos 2\phi_q \times  \nn \\
&& \langle 0_M |    a_M(\bq,-) ( \cos \phi_q - i  \sin \phi_q  e^{2 i E_q t} B_{M+}(\bq) ) 
l_\bq 
 ( \cos \phi_q + i  \sin \phi_q  e^{-2 i E_q t} B^\dagger_{M+}(\bq) )
a^\dagger_M(\bq,-)  |0_M \rangle .\nn \\
\label{eq:matrixelement}
\eea
The last matrix element in Eq.(\ref{eq:matrixelement}) is computed as;
\bea
&&  N_q^2  \cos 2\phi_q  \times \nn \\
&&
\langle 0_M |    a_M(\bq,-) ( \cos \phi_q - i  \sin \phi_q  e^{2 i E_q t} B_{M+}(\bq) ) 
l_\bq 
 ( \cos \phi_q + i  \sin \phi_q  e^{-2 i E_q t} B^\dagger_{M+}(\bq) )
a^\dagger_M(\bq,-)  |0_M \rangle \nn \\
&=&N_q^2 \cos 2 \phi_q  [\cos^2 \phi_q \langle 0_M |    a_M(\bq,-) l_\bq  a_M^\dagger(\bq,-) |0_M \rangle \nn \\
&&+\sin^2 \phi_q \langle 0_M |    a_M(\bq,-) B_{M+}(\bq)  l_\bq   B^\dagger_{M+}(\bq)  a_M^\dagger(\bq,-) |0_M \rangle \nn \\
&&-i  \sin \phi_q \cos \phi_q e^{2 i E_q t}  \langle 0_M | a_M(\bq,-) B_{M+}(\bq)   l_\bq   a_M^\dagger(\bq,-) |0_M \rangle  \nn \\
&& +i \sin \phi_q \cos \phi_q e^{-2 i E_q t}  \langle 0_M | a_M(\bq,-) l_\bq   B^\dagger_{M+}(\bq)   a_M^\dagger(\bq,-) |0_M \rangle 
]\nn \\
&=& N_q^2 \cos 2 \phi_q \Biggl{[}\frac{|\bq|}{E_q}\cos2\phi_q+\frac{m}{E_q}  \sin2\phi_\bq \cos (2 E_q t)\Biggr{]} (2\pi)^3 2 E_q \delta^{(3)}(0) \nn \\
&=& \left(\frac{q}{E_q}\right)^2+\left( \frac{m}{E_q} \right)^2 \cos (2E_q t),
 \eea
and for $l_\bp$ ($\bp \ne \bq$), the matrix element vanishes as;
\bea
 && \langle \Psi(t) |l_\bp|\Psi(t) \rangle\nn \\
&& = \prod_{\bk (\ne \bp) \in A}  \langle 0_M | 
 \{ \cos^2\phi_k-\sin^2\phi_k e^{4 i E_k t} 
  B_{M -}(\bk) B_{M +}(\bk)
 -i \sin \phi_k \cos \phi_k    e^{i 2 E_k t} \sum_{\lambda=\pm 1} B_{M \lambda}(\bk) 
 \} \nn \\
&&  \{ \cos^2\phi_k-\sin^2\phi_k e^{-4 i E_k t} 
B^\dagger_{M +}(\bk)  B^\dagger_{M -}(\bk) 
 +i \sin \phi_k \cos \phi_k    e^{-i 2 E_k t} 
\sum_{\lambda=\pm 1} B^\dagger_{M \lambda}(\bk) 
 \}  |0_M \rangle \nn \\
&& \times \langle 0_M | 
 \{ \cos^2\phi_p-\sin^2\phi_p e^{4 i E_p t} 
  B_{M -}(\bp) B_{M +}(\bp)
 -i \sin \phi_p \cos \phi_p    e^{i 2 E_p t} \sum_{\lambda=\pm 1} B_{M \lambda}(\bp) 
 \} |l_\bp \nn \\
&&  \{ \cos^2\phi_p-\sin^2\phi_p e^{-4 i E_p t} 
B^\dagger_{M +}(\bp)  B^\dagger_{M -}(\bp) 
 +i \sin \phi_p \cos \phi_p    e^{-i 2 E_p t} 
\sum_{\lambda=\pm 1} B^\dagger_{M \lambda}(\bp) 
 \}  |0_M \rangle  
\nn \\
 &&\times  N_q^2  \cos 2\phi_q  \times \nn \\
&&  \langle 0_M |    a_M(\bq,-) ( \cos \phi_q - i  \sin \phi_q  e^{2 i E_q t} B_{M+}(\bq) ) 
 ( \cos \phi_q + i  \sin \phi_q  e^{-2 i E_q t} B^\dagger_{M+}(\bq) )
a^\dagger_M(\bq,-)  |0_M \rangle 
\nn \\
&&= 
 \langle 0_M | 
 \{ \cos^2\phi_p-\sin^2\phi_p e^{4 i E_p t} 
  B_{M -}(\bp) B_{M +}(\bp)
 -i \sin \phi_p \cos \phi_p    e^{i 2 E_p t} \sum_{\lambda=\pm 1} B_{M \lambda}(\bp) 
 \} |l_\bp \nn \\
&&  \{ \cos^2\phi_p-\sin^2\phi_p e^{-4 i E_p t} 
B^\dagger_{M +}(\bp)  B^\dagger_{M -}(\bp) 
 +i \sin \phi_p \cos \phi_p    e^{-i 2 E_p t} 
\sum_{\lambda=\pm 1} B^\dagger_{M \lambda}(\bp) 
 \}  |0_M \rangle\nn \\
&& =0. 
\eea
Using the above results, one obtains the expectation value Eq.(\ref{eq:expextationvalue}) as;
\bea
\langle \Psi(t)|L|\Psi(t) \rangle&=& \left(\frac{q}{E_q}\right)^2+
\left( \frac{m}{E_q} \right)^2 \cos (2E_q t), \nn \\
&=& v^2+(1-v^2) 
\cos ( \frac{2m t}{\sqrt{1-v^2}}),
\label{eq:leptonnum}
\eea
with the velocity $v$ is defined by $v=\frac{q}{E_q}$.
In Fig.\ref{fig1}, Eq.(\ref{eq:leptonnum}) is plotted for various velocities of the neutrino.
For a relativistic neutrino $v \sim 1$, the lepton number stays around 1 with short period of the oscillation. For a non-relativistic neutrino $v < 1$, the lepton number oscillates between $1$ and $-1$ with the longer period. The period of the oscillation $T$ satisfies,
\bea
mT=\pi \sqrt{1-v^2}.
\eea
\section{Conclusions}
We study the time evolution of lepton number in the Schr\"{o}dinger picture for a single
flavor.  Similar to the Heisenberg picture, a multi-flavor   Schr\"{o}dinger is also possible.
But the extention to the multi-flavor case is outside of the scope of the paper.
The expectation value is the same as that obtained by Heisenberg picture previously.
For non-relativistic case, it oscillates with the large amplitude ($\langle \Delta L \rangle \simeq \pm1$). 
For relativistic case, it oscillates rapidly with the small amplitude around  $\langle L \rangle \simeq 1$. 
The vacuum with null lepton number is a superposition of the vacuum for mass eigenstate, a pair of
Majorana particles, and two pairs of Majorana particles. 
Similar to the vacuum, the one particle state with the definite lepton number $1$  is a  
superposition of a mass eigenstate of Majorana neutrino and a state with a Majorana pair and a Majorana  neutrino. 
These non-trivial superposition of states with different energies give rise to the oscillating behavior for the expectation value of lepton number. \\

\vspace{1cm}
\noindent
{\bf Acknowledgement}
We would like to thank the organizers of the Corfu 2021. The work of T.M. is supported by Japan Society for the Promotion of Science (JSPS) KAKENHI Grant Number JP17K05418.

\appendix
\section{The derivation of  Eqs.(\ref{eq:Bogo1}-\ref{eq:Bogo4}) and  Eq.(\ref{eq:Tp}) }
In this appendix, we prove the relations Eqs.(\ref{eq:Bogo1}-\ref{eq:Bogo4}).
\bea
&& T(\phi_p) a^\dagger_M(\bp,-)  T^{-1}(\phi_p) =\cos \phi_p a_M^\dagger (\bp, -) - i \sin \phi_p a_M (-\bp, -) , \\
&& T(\phi_p) a^\dagger_M(-\bp,-)  T^{-1}(\phi_p) =\cos \phi_p a_M^\dagger (-\bp, -) + i \sin \phi_p a_M (\bp, -),
\eea
with  the definition of  $T(\phi_p)$ in
 Eq.(\ref{Transform}) and Eq.(\ref{eq:Tphip}). One can also compute derivatives,
\bea
&& \frac{d^{n}}{d\phi_p^{n}} \{T(\phi_p) a^\dagger_M(\pm \bp,-)  T^{-1}(\phi_p)\} \nn \\
&=& i^{n} [g_{p,+}+g_{p,-},[g_{p,+}+g_{p,-}, ... [g_{p,+}+g_{p,-}, \{T(\phi_p) a^\dagger_M(\pm \bp,-)  T^{-1}(\phi_p)\}] .
\eea
For $n=2m$($m=$ integer), it is 
\bea
&& \frac{d^{2m}}{d\phi_p^{2m}} \{T(\phi_p) a^\dagger_M(\pm \bp,-)  T^{-1}(\phi_p)\}\Bigr{|}_{\phi_p=0}=(-1)^{m} a^\dagger_M(\pm \bp,-).
\eea
For $n=2m+1$, it is given by,
\bea
&& \frac{d^{2m+1}}{d\phi_p^{2m+1}} \{T(\phi_p) a^\dagger_M(\pm \bp,-)  T^{-1}(\phi_p)\}\Bigr{|}_{\phi_p=0}=(-1)^{m} (\mp i) a_M(\mp \bp,-).
\eea
They are derived with the following commutation relations,
\bea
&& [g_{p,+}+g_{p,-}, a^\dagger_M(\pm \bp,-)]=\mp a_M(\mp \bp,-). 
\eea
Then one can show,
\bea
T(\phi_p) a^\dagger_M(\bp,-)  T^{-1}(\phi_p)&&=\left(\sum_{m=0}^{\infty} \frac{(-1)^m}{(2m)!} (\phi_p)^m  \right) a^\dagger_M(\bp,-) -i \left(\sum_{m=0}^{\infty} \frac{(-1)^m}{(2m+1)!}  (\phi_p)^{m+1}  \right)a_M(-\bp,-) \nn \\
&&= \cos \phi_p   a^\dagger_M(\bp,-) -i \sin \phi_p a_M(-\bp,-),
\\ 
 T(\phi_p) a^\dagger_M(-\bp,-)  T^{-1}(\phi_p)&& = \cos \phi_p a^\dagger_M(-\bp,-)  +i \sin \phi_p a_M(\bp,-). 
\eea
This leads to the relations of Eq.(\ref{eq:Bogo1}) and Eq.(\ref{eq:Bogo2}).
Eqs.(\ref{eq:Bogo3}-\ref{eq:Bogo4}) can also be proved similarly.
Next, the outline of the derivation of Eq.(\ref{eq:Tp}) is given as follows;
\bea
T(\phi_p) |0_M \rangle&=&\exp[+i \sum_{\lambda=\pm} g_\lambda(p) \phi_p]|0_M \rangle \nn \\
&=&\sum_{n=0}^{\infty} \frac{i^n (\phi_p)^n }{n!} (g_+(p)+g_-(p))^n  |0_M \rangle , \nn  
\eea
where we denote $g_\lambda(p)=g_{p, \lambda}$.
\bea
 (g_+(p)+g_-(p)) |0_M \rangle&=& \sum_{\lambda=\pm} B^\dagger
_{M \lambda}(\bp) |0_M \rangle , \\
\{ g_+(p)+g_-(p) \}^2|0_M \rangle&=&2(1+B^\dagger_{M -}(\bp)B^\dagger_{M +}(\bp))|0_M \rangle, \\
\{ g_+(p)+g_-(p) \}^3|0_M \rangle&=&2 \{ g_+(p)+g_-(p) \}(1+B^\dagger_{M -}(\bp)B^\dagger_{M +}(\bp))|0_M \rangle \nn \\
&=& 4 \{ g_+(p)+g_-(p) \}|0_M \rangle\nn \\
&=&4 \sum_{\lambda=\pm} B^\dagger
_{M \lambda}(\bp) |0_M \rangle ,\label{eq:g3} \\
 \{ g_+(p)+g_-(p) \}^4|0_M \rangle &=&4\{ g_+(p)+g_-(p) \}^2 |0_M \rangle.
\eea
In deriving Eq.(\ref{eq:g3}), we have used the following anti-commutation relation,
\bea
\{B_{M \lambda}(\bp), B^\dagger_{M \lambda}(\bp) \} &=&
\alpha_M(\bp,\lambda) \alpha^\dagger_M(\bp,\lambda)+
\alpha^\dagger_M(-\bp,\lambda) \alpha_M(-\bp,\lambda) \nn \\
&=&1-\alpha^\dagger_M(\bp,\lambda) \alpha_M(\bp,\lambda)+\alpha^\dagger_M(-\bp,\lambda) \alpha_M(-\bp,\lambda).
\eea
From the consideration above, one concludes that 
\bea
&&\{ g_+(p)+g_-(p) \}^{2n+1}|0_M \rangle= 4^n \{ g_+(p)+g_-(p) \}|0_M \rangle,\nn \\
&&\{ g_+(p)+g_-(p) \}^{2n+2}|0_M \rangle=4^n \{ g_+(p)+g_-(p) \}^2|0_M \rangle.
\eea
Then  one obtains the relation in Eq.(\ref{eq:Tp}), 
\bea
&& T(\phi_p) |0_M \rangle \nn \\
&=& |0_M\rangle + i \sum_{n=0}^{\infty}\frac{(-1)^n \phi_p^{2n+1} 2^{2n}}{(2n+1)!}  \{ g_+(p)+g_-(p) \}|0_M \rangle
+\sum_{n=0}^{\infty}\frac{(-1)^{n+1} \phi_p^{2n+2} 2^{2n}}{(2n+2)!}  \{ g_+(p)+g_-(p) \}^2|0_M \rangle \nn \\
&=& |0_M\rangle -\frac{\sin^2 \phi_p}{2} \{ g_+(p)+g_-(p) \}^2|0_M \rangle+ i \cos \phi_p \sin \phi_p  \{ g_+(p)+g_-(p) \}|0_M \rangle \nn \\
&=& \cos^2\phi_p |0_M\rangle + i \cos \phi_p \sin \phi_p \sum_{\lambda=\pm} B^\dagger
_{M \lambda}(\bp) |0_M \rangle -\sin^2 \phi_p B^\dagger_{M -}(\bp)B^\dagger_{M +}(\bp)|0_M \rangle,
\label{eq:ATp}
\eea
where we used the following relations
\bea
\sum_{n=0}^{\infty}\frac{(-1)^n \phi_p^{2n+1} 2^{2n}}{(2n+1)!}= -\frac{1}{2} \sin^2 \phi_p, \quad
\sum_{n=0}^{\infty}\frac{(-1)^{n+1} \phi_p^{2n+2} 2^{2n}}{(2n+2)!}=
 \cos \phi_p \sin \phi_p.
\eea
\noindent
\section{The derivation of Eq.(\ref{eq:initial})}
In this appendix, we derive the relation which is used to derive Eq.(\ref{eq:initial}),
\bea
 &&\sqrt{\cos 2\phi_q} (\cos \phi_q a^\dagger_M(\bq,-) - i 
\sin \phi_q a_M(-\bq,-) )  \nn \\
&& \times
\{ \cos^2\phi_q-\sin^2\phi_q 
B^\dagger_{M +}(\bq)  B^\dagger_{M -}(\bq) 
 +i \sin \phi_q \cos \phi_q  \sum_{\lambda=\pm} B^\dagger_{M \lambda}(\bq) 
 \} |0_M \rangle \nn \\ 
&=& \sqrt{\cos 2\phi_q}\{ (\cos^3 \phi_q+\sin^2\phi_q
\cos \phi_q) a_M^\dagger(\bq,-) \nn \\
&& + i (\sin \phi_q \cos^2 \phi_q +\sin^3 \phi_q)   B^\dagger_{M +}(\bq) a_M^\dagger(\bq,-) \}|0_M \rangle \nn \\
&=&  \sqrt{\cos 2\phi_q}\{ 
\cos \phi_q a_M^\dagger(\bq,-) + i  \sin \phi_q  B^\dagger_{M +}(\bq) a_M^\dagger(\bq,-) \} |0_M \rangle.  
\label{eq:B1}
\eea

\end{document}